\documentstyle[12pt]{article}
\input psfig
\long\def\comment#1{}
\begin{document}
\title{General  Qubit Errors Cannot Be Corrected}
\author{Subhash Kak}
\date{June27, 2002}
\maketitle

\begin{abstract}
Error correction in the standard meaning of the term
implies the ability to correct all small analog errors
and some large errors.
Examining assumptions at the basis of
the recently proposed quantum error-correcting codes,
it is pointed out that these codes can correct only
a subset of errors, and are unable to
correct small phase errors which can have disastrous 
consequences for a quantum computation. This shortcomings
will restrict their
usefulness in real applications.

\end{abstract}

\thispagestyle{empty}
\begin{quote}
{\small {\it Colours seen by candle-light\\
Will not look the same by day.}\\
	\hspace*{2in}		-Robert Browning}
\end{quote}

\section{Introduction}

Since the work of Calderbank, Shor, and Steane\cite{Ca95,St96,Go00} (CSS),
the question of error-correction coding for quantum computing
has attracted much attention and several codes have
been proposed.
But these codes have been devised to work under very restrictive conditions and
they can potentially correct only bit flips and phase flips
and some combinations thereof,
which errors represent a small
subset of all the errors that can corrupt a quantum state.
This would not be an issue if phase errors were not important in
a quantum computation. But they are, since we manipulate the phases
to drive a quantum computation to a useful conclusion.
Many quantum algorithms require the computation begin
with no phase errors in the start of the computation.
CSS codes cannot correct for errors such as a
$\frac{1}{\sqrt 2} ( | 0\rangle + | 1\rangle$ changing into a
$\frac{1}{\sqrt 2} ( | 0\rangle + e^{i 0.002} | 1\rangle$,
without assuming that there is some part of the code
that does not suffer any error at all, no matter
how small.

It might appear odd for anyone to question quantum error 
correction when researchers have been working in this area
for several years. 
Actually, the ability of
quantum error correction methods to eliminate
analog quantum noise has been questioned before\cite{Ka99}.
Analog errors in the analog domain (which is like the
quantum information situation) can simply not be
completely eliminated. Using redundancy could, at best,
reduce errors under appropriate conditions.
Therefore, the claim that analog quantum errors could somehow be
eliminated  has been found puzzling, especially because
the No-Cloning theorem makes it impossible to copy
quantum states.
But the proponents of quantum
error correction codes  felt that ``quantum errors could be
fixed with quantum tricks\cite{Go00}.''
These tricks seem to work because
the term ``error correction''
in quantum computing has been used in a non-standard manner.
But we wish to stress that this is not just a semantic
problem.

Error correction, intuitively and in classical theory,
implies that if
\[ y = x + n,\]
where $x$ is the discrete codeword, $n$ is analog noise, and
$y$ is the analog noisy codeword, one can 
recover $x$  {\it completely and fully} so long as the
analog noise function $n$ is less than a certain 
threshold.
If it exceeds this threshold, then also there is full 
correction so long as this does not happen more than
a certain number of times (the
Hamming distance for which the code is designed) at the 
places the analog signal $y$ is sampled.

In other words, the hallmark of
classical  error-correcting codes is {\it the correction of 
all possible small analog errors} and many others which
exceed the thresholds associated with the code alphabet.
This full correction of all possible small analog errors is
beyond the capability of the proposed quantum error-correcting codes.

This definition of error correction in classical theory
is not merely a matter of convention or intuition.
In classical information science, errors are
analog and, therefore, all the possible small errors
must be corrected by error-correcting codes.
To someone who looks at this field from the outside, it might
appear that one only needs to fix bit flips.
In reality, small analog errors, occurring on
all the bits, are first removed by the use
of clamping and hard-limiting.

Since the definition of a qubit includes arbitrary phase, 
it is necessary to consider errors from the perspective
of the quantum state and not just from that of final measurement.
As mentioned above, in the classical theory, it is implicitly accepted
that all possible small analog errors have already been
corrected by means of an appropriate  thresholding operation.
Therefore, we must define correction of small analog
phase errors as a requirement for quantum error correction.

This paper reviews assumptions behind the CSS quantum
error-correcting codes.
The construction of these codes requires precise knowledge
of 
the state of the coded qubit, in which no phase 
uncertainties are conceded.
This precisely known coded qubit state helps to determine
a standard
against which errors are measured.
This precision will not be available in practice.

The paper is organized as follows:
Section 2 briefly reviews the CSS model,
Section 3 presents the qubit sphere to highlight
the difficulty posed by unknown phase, and
Section 4 considers the question of what errors
can be corrected, which is followed by conclusions.
\section{The quantum error correction model}

A quantum error-correcting code is defined to be a unitary
mapping of $k$ qubits into a subspace of the quantum state
space of $n$ qubits such that if any $t$ of the qubits 
undergo arbitrary decoherence, not necessarily independently,
the resulting $n$ qubits can be used to faithfully reconstruct
the original quantum state of the $k$ encoded qubits\cite{Ca95}.
The assumptions in the quantum error correction model are\cite{St96}:
Arbitrary errors of qubits are divided into 
`amplitude errors', that is, changes of the form
$|0\rangle \leftrightarrow |1\rangle$, and `phase errors',
that is, changes of the form
$|0\rangle + |1\rangle \leftrightarrow |0\rangle - |1\rangle$.

These assumptions seem to have been made with the final
measurement in mind, where the objective is to 
get a binary sequence from the measurement apparatus.
The idea here is that if $0$s have been converted into
$1$s and vice versa, the redundancy of the error-correction
code will be able to tell us where the error has occurred,
allowing us to reconstruct the correct sequence.

A quantum system is correctly viewed as being apart from the
observer, who enters the picture only when the measurement is
made. This means that one can speak of two perspectives as
far as errors are concerned: (A), errors relative to the
quantum state itself; and (B), errors relative to the
observer who will make the measurement. Since the
transformation between the quantum state and the
measurement is {\it many-to-one}, the two perspectives
are not identical.
The CSS model considers the second perspective only, without
relating it to the errors in the quantum state.
By doing so, the model misses out on errors that can have
a catastrophic effect on the computing process.

Note that classical error correction theory does not 
bother about such a dual perspective because of two
reasons: first,
the absence of
anything analogous to state collapse; second, the small
analog errors are assumed to have been corrected by
a hard-limiting operation prior to converting the
received analog $y$ signal by sampling into the
discrete, binary codeword.
In classical theory, all the useful information within the system
is accessible, which is not the case in a quantum system.

The perspective B is described elsewhere by the 
author\cite{Ka99,Ka00,Ka01a,Ka02}, where it
is argued that random, small errors in phase as well
as admixture of unwanted states can be
problematic for the implementation of quantum algorithms.

\paragraph{Phase errors in the codeword}

In one well known one qubit error-correcting
code, each qubit is represented by seven
qubits.
The seven qubit system is interpreted as a 
pair of abstract particles: the abstract
qubit, and the syndrome space.
The idea behind the method is that the error will
leave the state component unchanged, and by measuring
the syndrome one would know the unitary transformation
to be applied to correct the error.
The code for $| 0 \rangle$ has an even number of
$1$s and the code for $|1 \rangle$ has an odd number
of $1$s. In reality, the coded qubits should be:

\[| 0\rangle_{code} = \frac{1}{\sqrt8} (|0000000\rangle + e^{i \theta_{01}} |0001111\rangle+ e^{i \theta_{02}}
|0110011\rangle + e^{i \theta_{03}} |0111100\rangle\\\]

\begin{equation}
 + e^{i \theta_{04}} |1010101\rangle
+e^{i \theta_{05}} |1011010\rangle +e^{i \theta_{06}} | 1100110\rangle + e^{i \theta_{07}} |1101001\rangle),
\end{equation}

\[|1\rangle_{code} = \frac{1}{\sqrt8} (|1111111\rangle + e^{i \theta_{11}} |1110000\rangle+e^{i \theta_{12}} 
|1001100\rangle + e^{i \theta_{13}} |1000011\rangle \\\]

\begin{equation}
+ e^{i \theta_{14}} |0101010\rangle
+e^{i \theta_{15}} |0100101\rangle +e^{i \theta_{16}} | 0011001\rangle +e^{i \theta_{17}}  |0010110\rangle).
\end{equation}
where $\theta_{ij}$ are random phase errors.
But in the theory, the uncertainties related to $\theta_{ij}$ 
are taken to be zero.
This makes it possible to use the codewords as the standard against
which other errors can be checked. In a realistic theory
the $\theta_{ij}$ cannot be taken to be zero.

Similarly, in the 9-cubit code, the codewords should be:

\begin{equation}
|0\rangle_{code} =  (|000\rangle + e^{i \theta_{1}} |111\rangle)
(|000\rangle + e^{i \theta_{2}} |111\rangle)
(|000\rangle + e^{i \theta_{3}} |111\rangle)
\end{equation}

\begin{equation}
|1\rangle_{code} =  (|000\rangle - e^{i \theta_{4}} |111\rangle)
(|000\rangle - e^{i \theta_{5}} |111\rangle)
(|000\rangle - e^{i \theta_{6}} |111\rangle)
\end{equation}
where $\theta_i$ are small phase errors.
But, again,  it is assumed that the $\theta{i}$s are zero.

Computing the overlap of the codewords (3) and (4) with
error states shows clearly that there would be a large
probability that small errors will not be corrected.
Neither the small random phase errors in the codewords, 
nor those that occur later,
will be eliminated.

\paragraph{Ancilla qubits}
We refer the reader to the CSS constructions where ancilla bits
are used to obtain the noise free state of the quantum 
code\cite{St97,Go00}.
The ancilla bits are assumed to be in the precise
all zero state, with no phase errors, whatsoever!
Steane acknowledges\cite{St97}\footnote{Page 39} that for quantum error
correction to work the assumption that the ancilla be
noise free needs to be dropped.
He suggests that fault-tolerant quantum computation\cite{Sh96}
will help alleviate this difficulty.
But the fault-tolerant system only shifts the burden
by assuming zero phase errors elsewhere in the
constructions.

Even for the correction of a single qubit, there is
circularity of argument.
One needs perfect ancilla bits, and {\it even if} we had
them, one can allow for only one error in a 9-qubit code.
How can one guarantee that there will be {\it absolutely
no phase error} -- no matter how small -- in the rest of the 9
qubits?

\section{The qubit sphere}

To examine the perspective A, which is with respect to the
quantum state,
it's useful to
begin with the representation of a qubit 
as the superposition
$| \phi\rangle = \alpha e^{i \theta_1} | 0\rangle + \beta  e^{i \theta_2}| 1\rangle,$
where $ \alpha, \beta \in R$ and $\alpha^2 + \beta^2 = 1$,
as a 
four-dimensional sphere.
To simplify matters, we consider only the difference
in phases and reduce the qubit to
$| \phi\rangle = \alpha | 0\rangle + \beta  e^{i \theta}| 1\rangle.$
The qubit is now a triple $(\alpha, \beta, \theta)$ and it
can be represented by a three-dimensional sphere of
 Figure 1.

\vspace{2mm}
\begin{figure}
\hspace*{0.5in}\centering{
\psfig{file=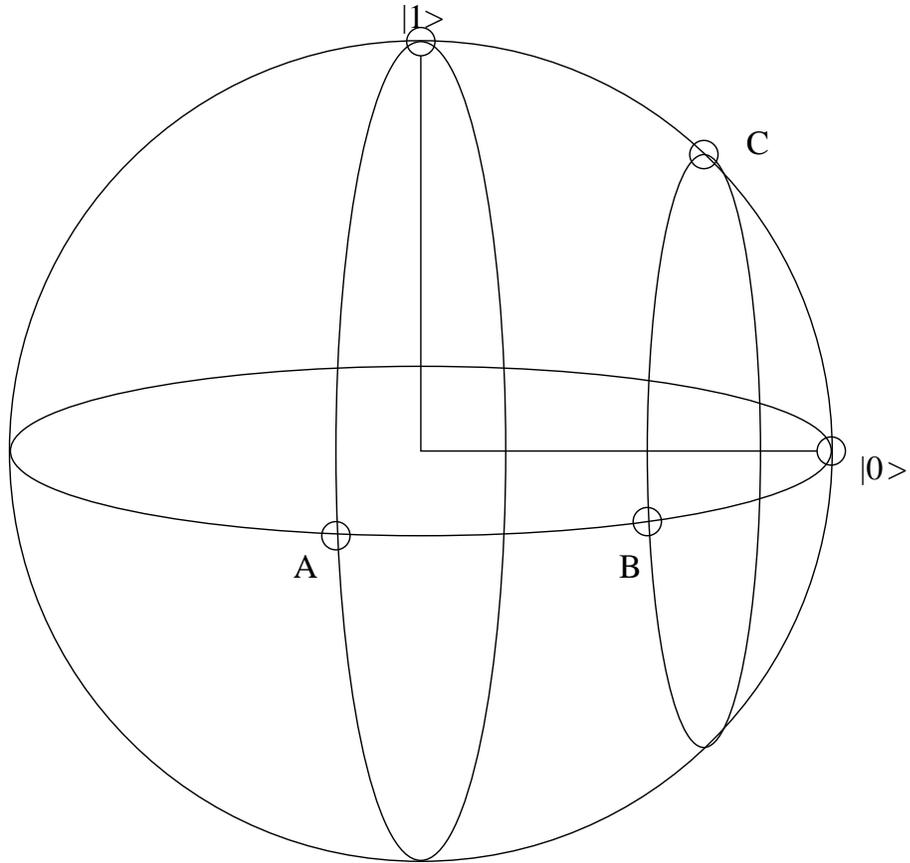,width=12cm}}
\caption{The qubit sphere $( \alpha, \beta, \theta)$.
The vertical circles represent $|1\rangle$ and its phase
shifts. 
The circle on the right represents 
$ 1/2^{1/2}(|0\rangle + e^{i \theta} |1\rangle ) $, which are
various combinations of $|0\rangle$ with phase shifted $|1\rangle$
(i.e. $45^o$ polarized photons, for example).
The point A is
$  e^{i \pi /2} | 1\rangle$; B is
$ 1/2^{1/2}(|0\rangle + i |1\rangle ) $; C is
$ 1/2^{1/2}(|0\rangle +  |1\rangle ) $.
}
\end{figure}

\vspace{2mm}

Parenthetically, let it be noted that
our qubit sphere 
is drawn differently from the qubit sphere of Tittel and Weihs\cite{Ti01},
who show $|0\rangle$ and $|1\rangle$ as opposite points on the same
circle on the sphere.

In the qubit sphere of Figure 1, the motion counterclockwise is taken to
be positive. The point of intersection of the two spheres at the front
end will be the state
$i | 1\rangle$.

Assuming, for example, that we are speaking of polarized photons,
we see that with respect to $|0\rangle$ the $45^o$ polarized photons
are points anywhere on the circle to the right.
Also, if there is unknown phase associated with $|0\rangle$, the
$45^o$ photons can be anywhere on the sphere surface\cite{Ka00}.

CSS considers just four points
$|0\rangle$, $|1\rangle$, and their sums and differences
on the qubit sphere, because doing this reduces the
quantum problem to two separate classes of classical
error correction.
These four points represent a small subset of
all the points on the qubit sphere.
Furthermore, the location of these four points will
be characterized by small errors.

\section{What errors can be corrected?}

Error correction is possible only for discrete quantities.
In classical information theory, error correction of 
a single bit is possible because there is a separation in
amplitude between 0 and 1. 
When bit flips between these two values
are considered, one can, by introducing
redundancy, increase distance between codewords,
ensuring the capacity to correct certain errors.
The CSS method appears to do the same thing ensuring that
under the assumed noise model
it will work fine as long as the qubits suffer only bit and
phase flips and their combinations.
But these errors are a small subset of all the errors that
are possible.

The idea of using bit flips and phase flips comes from the
fact that the
Pauli group consists of four operators: identity ($I$), bit flip
($X$),
phase flip ($Z$), and bit-and-phase flip ($Y$).
These four matrices:

$I = 
\left( \begin{array}{cc}
      1  & 0 \\
	0 & 1\\
                               \end{array} \right)$,
$X = 
\left( \begin{array}{cc}
      0  & 1 \\
	1 & 0\\
                               \end{array} \right)$,
$Z = 
\left( \begin{array}{cc}
      1  & 0 \\
	0 & -1\\
                               \end{array} \right)$,
$Y = 
\left( \begin{array}{cc}
      0  & -i \\
	i & 0\\
                               \end{array} \right)$\\
span the space of $2\times 2$ matrices, and the n-qubit
Pauli group spans the space of $2^n \times 2^n$
matrices. A general phase error will then be represented
by a linear combination of bit and phase errors.

However, to correct an analog phase error one still needs 
perfectly error-free ancilla qubits which is impossible to
guarantee and, therefore, we are
unable to proceed further.

The CSS noise model is restrictive
from a
practical point of view. It ignores
that each qubit,
being a triple $(\alpha, \beta, \theta)$,  will have small, unknown values
initially,
even when the strategy of using atom cooling is employed to
generate a coherent state.

Furthermore, the application of quantum algorithms 
by means of electric and magnetic fields, and decoherence, will introduce 
additional
phase uncertainty.
Small phase errors will become large as unitary transformations are
applied repeatedly in the execution of a quantum algorithm.
Since quantum calculations 
are sensitive to the phase values, they will have
uncontrollable effects.

In fact, the starting
states will not only have small random phase errors,
but also an admixture of all other states, albeit
with small complex amplitudes.
This introduces an additional complicating factor which the
CSS model ignores.

Just as classical error models assume the same type
of analog error corrupting each bit, one needs to
accept that analog error will corrupt {\it each} 
qubit.
But an
analysis of such a situation, given further that
the initial state is correctly seen as
an admixture, will be difficult.
As a start,
it may be useful to determine the influence on performance
of random phase errors in the qubit state and those
in the measurement of the syndrome state.

Only discrete quantities to which small values of noise are
added can be corrected;
noise added to  an analog variable cannot be removed, and
quantum phase is an analog variable.
Analog quantities (such as
qubit phases) cannot be corrected unconditionally.

One can measure analog variables with respect to a
standard, and then correct any deviations from the
standard. This is what appears to be happening in the
disregarding of
random phases in the coded qubit. But 
that is tantamount to a backdoor 
discretization of the problem.

\section{Conclusions}
Error correction requires correction of all small 
errors and some large errors.
This the CSS
quantum error correction model 
is unable to do.

The error model used by CSS is not realistic. It assumes 
zero phase errors in many of its constructions,
which precision will be absent in 
the real world.
There can never be any guarantee of
zero phase error in the qubits.
Unlike classical error models where each bit
is corrupted by noise, the CSS model assumes that
most qubits are perfectly precise.
This only shifts the task from error correction
to initialization, without indicating how that 
might be done\cite{Ka02}. 
As far as 
the constructions of quantum error correcting codes refer to
physical reality, they are not certain, and as far as they are certain,
they do not refer to physical reality.
A realistic error model must assume that all qubits,
including the ancilla bits, have the same type
of errors.

Because qubits are arbitrary combinations of
$|0\rangle$s and $|1\rangle$s
$(\alpha, \beta, \theta)$,
lack of knowledge of the relative phase
can send the qubit to any point on the sphere.
The CSS model is a less than successful joining
of the classical error-control theory to quantum information.
It violates the basic premise of
error correction, that {\it it should be possible to
correct all possible small errors}, and some large errors.

The CSS  model may be called a method of error reduction,
under narrow conditions of some of the qubits escaping
all error.
But if there are perfect, error-free qubits, why not 
use them in the first place?
Error reduction, even if there was a
way of estimating it in the presence of unknown
small phase errors, 
may  not be of use in quantum
computing techniques where absolutely no error 
is
permitted for useful computation to take place.

We cannot be hopeful for other methods of 
qubit error correction either, since the
difficulty arises out of the analog
nature of the error process.

\paragraph{Acknowledgement}
I would like to thank Daniel Gottesman
and Andreas Klappenecker for their comments on
an earlier version of this paper.
I am especially grateful to
Michel Dyakonov
for criticism and sage advice.

\end{document}